\documentclass[prl,twocolumn,superscriptaddress,showpacs]{revtex4}
\usepackage{times}
\usepackage{amsmath}
\usepackage{amssymb}
\usepackage{epsfig}
\usepackage{graphicx}
\usepackage{wasysym}

\newcommand \be{\begin{equation}}
\newcommand \ee{\end{equation}}
\newcommand \bes{\begin{equation*}} 
\newcommand \ees{\end{equation*}}
\newcommand \bea{\begin{eqnarray}}
\newcommand \eea{\end{eqnarray}}
\newcommand \beas{\begin{eqnarray*}} 
\newcommand \eeas{\end{eqnarray*}}
\newcommand \bfg{\begin{figure}}
\newcommand \efg{\end{figure}}
\newcommand \bfgs{\begin{figure*}} 
\newcommand \efgs{\end{figure*}}
\newcommand \bwt{\begin{widetext}}
\newcommand \ewt{\end{widetext}}
\newcommand \dif{{\rm d}}
\newcommand \im{i}

\newcommand \nd{{\vphantom{\dagger}}} 
\newcommand \bra{\langle}
\newcommand \ket{\rangle}
\newcommand \fig[1]{FIG.~\ref{#1}}
\newcommand \eqtn[1]{(\ref{#1})}
\newcommand \tbl[1]{Table~\ref{#1}}

\newcommand \inlinecite[1]{Ref.~\onlinecite{#1}}

\newcommand \ising{\mathcal{S}}
\newcommand \dimer{\mathcal{C}}

\newcommand \modified{}

\begin{document}
\title{Extended supersolid phase of frustrated hard-core bosons on a triangular lattice}
\author{Fa Wang, Frank Pollmann, Ashvin Vishwanath}
\affiliation{Department of Physics, University of California, Berkeley, CA94720}
\date{\today}

\begin{abstract}
We study a model of hard-core bosons with {\em frustrated}
nearest-neighbor hopping ($t$) and repulsion ($V$) on the triangular
lattice.
We argue for a supersolid ground state in the large repulsion
($V\gg|t|$) limit where a dimer representation applies, by
constructing a unitary mapping to the well understood unfrustrated
hopping case. This generalized 'Marshall sign rule' allows us  to
establish the precise nature of the supersolid order by utilizing a
recently proposed dimer variational wavefunction, whose correlations
can be efficiently calculated using the Grassman approach,.
By continuity, a supersolid is predicted over the wide parameter
range, $V>-2t>0$.
This also establishes a simple phase diagram for the triangular
lattice spin 1/2 XXZ antiferromagnet.
\end{abstract}
\maketitle

Supersolidity, where superfluid and crystalline orders coexist, have
fascinated physicsits since they were first theoretically
proposed\cite{Andreev69}. Recent experimental results in
$^4$He\cite{MosesChan} that are still under active debate have led
to renewed interest. Experimental developments on a different front,
in the realization of optical lattices in ultracold atomic systems,
motivated a search for a lattice supersolid. One of the more
promising candidates is a model of strongly interaction hard-core
bosons on a triangular lattice. The model Hamiltonian reads \be
 H=-t\sum_{\langle ij\rangle}\left(b_i^{\dag}b_j^{\vphantom{\dag}}+\text{H.c.}\right) + V\sum_{\langle ij\rangle}\left( n_i-\frac12\right)\left(n_j-\frac12 \right),
 \label{equ:Hbosons}
\ee where $b_i^{\vphantom{\dag}}$ ($b_i^{\dag}$) annihilates
(creates) a hard-core boson on site $i$ and
$n_i=b^{\dag}_ib_i^{\vphantom{\dag}}$ are density operators. The
model is equivalent to the XXZ
spin-1/2 Hamiltonian on the
triangular lattice
\be
 H=\sum_{\langle ij\rangle}\left[\frac{J_{\perp}}{2}
\left (s_i^+ s_j^- + s_i^- s_j^+\right)+ \frac{J_z}{4} s_i^z s_j^z\right],
 \label{equ:H}
\ee {\modified where $s^{\pm}=(1/2)(s^x\pm \im s^y)$, $s^{x,y,z}$
are the Pauli matrices and are related to bosons by
$s_i^z=(2n_i-1),\,s^+=b^\dagger,\,s^-=b^\nd$, and $J_z =V$,
$J_\perp=-2t$.} The discussion below will be largely in terms of the
bosons, although we will sometimes switch to the equivalent spin
description, when that is more natural.
The
$t>0$ case  corresponds to the unfrustrated hard-core boson model
with repulsive nearest-neighbor interactions.
For this case, a variety of studies including large scale quantum
Monte Carlo simulations \cite{Heidarian05, Melko05,Wessel05}
indicate a supersolid phase for all $V/t\ge8.9$, stabilized by an
`order by disorder' mechanism. The solid order is of the three
sublattice (++-) type, where two sublattices have the same boson
density.

\begin{figure}
 \includegraphics[width=7cm]{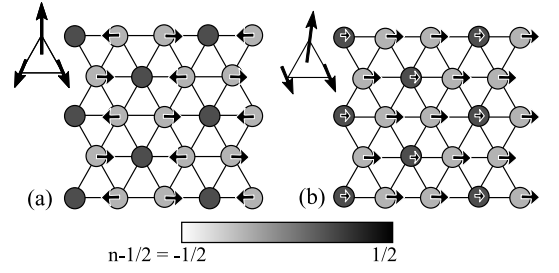}
\caption{Supersolid LRO from the variational wavefunction in the
strong repulsion limit. Greyscale shows density order $\langle
2n_i-1\rangle$ while arrows denote superfluid order $\langle
b_i^\dagger\rangle$, for (a): Frustrated hopping ($t<0$) - note the
sign structure of superfluid order; and (b):Unfrustrated hopping
($t>0$)}
\label{fig:order}
\end{figure}

The case of {\em frustrated} hopping ($t<0$) suffers from a sign
problem in the occupation number basis, and its ground state has
been a subject of conjecture for the last three decades. The
corresponding spin model is just the XXZ antiferromagnet, which,
in the large $J_z$ limit was at the center of the RVB spin liquid
proposal of Fazekas and Anderson \cite{FazekasAnderson}. Later
semiclassical and small cluster numerical studies suggested magnetic
order \cite{Fazekas}, and general arguments which apply to the phase
structure of bipartite dimer models, to which this model can be
mapped in the large $J_z$ limit, also indicate the same
result\cite{ReadSachdev,Fradkin}. However, the precise nature of
ordering has not been conclusively established. In this letter, we
show how this problem can be tackled,
which is summarized briefly in boson language below. Due to
frustration, the ground states in the $V\rightarrow \infty$ limit is
extensively degenerate.
Within this ground state manifold, we demonstrate that the
frustrated problem with $t<0$ can be mapped, via a nontrivial
unitary transformation, onto the unfrustrated one with $t>0$. Since
the latter is well understood\cite{Heidarian05,
Melko05,Wessel05,Prokofiev}, many properties of the frustrated case
can be immediately derived. Such a generalized `Marshall sign' was
conjectured earlier based on state enumeration and numerics
\cite{FazekasAnderson,ShengPrivate}. Here we construct the explicit
transformation which proves this conjecture, and moreover utilize it
to deduce properties of the frustrated model. Our unitary
transformation is diagonal in the occupation number basis, which,
combined with our knowledge of the unfrustrated model, allows us to
argue that a supersolid state is realized for $t<0$ as well. The
precise details of the superfluid phase ordering requires further
calculation. This is carried out using a variational wave-function
approach \cite{Sen08} recently introduced for the unfrustrated $t/V
= 0^+$ limit, which captures the essential aspects of supersolid
order very well and has good variational energy as compared to the
quantum Monte Carlo results. Applying the unitary transformation, we
obtain a variation wavefunction for the frustrated problem.
Properties of this wavefunction, in particular the phase
correlations, are then calculated. The state is found to be a
supersolid and the resulting structure of the long range order (LRO)
is shown in \fig{fig:order}a. Surprisingly, the superfluid amplitude
vanishes on one of the sublattices and hence superfluidity lives
exclusively on the honeycomb lattice formed by the remaining two
sublattices, on which the amplitude alternates in sign. Contrary to
naive expectations, the superfluid amplitude on these sites {\it
exceeds} the maximum superfluid amplitude of the unfrustrated case.

\begin{figure}
\includegraphics[scale=0.9]{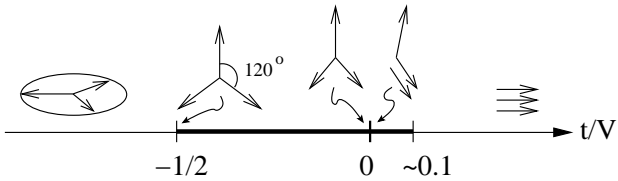}
\caption{ Schematic phase diagram for both unfrustrated ($t>0$) and
frustrated hopping ($t<0$) with repulsive interactions ($V>0$). The
three arrows are order parameters
 $\vec{s}=(b^\dagger+b,\im b-\im b^\dagger,2n-1)$
on the three sub-lattices. For $t/V < -1/2$ or $t/V > 0.1$ there is
only superfluid LRO (XY spin order). The thick line $-1/2 < t/V <
0.1$ is the region of supersolid order. $t/V=-1/2$ is the SU(2)
symmetric antiferromagnet.} \label{fig:phasediagramXXZ}
\end{figure}

Finally, with this information in hand, we propose a phase diagram
for the entire $t/V>0$ parameter range. Note the point
 $t/V=1/2$
corresponds to the spin-isotropic triangular antiferromagnet, where
the 120$^\circ$ state is established. This can be smoothly connected
to
the large $V$ supersolid state derived here as shown in
\fig{fig:phasediagramXXZ}. Supersolid order would then
naturally be preserved over the wide parameter range
 $0<-t<V/2$,
in contrast to the unfrustrated case, where it is only present for
$t<V/10$. The frustrated triangular lattice boson model therefore
appears to be an appealing candidate for the realization of the
elusive supersolid phase - experimental prospects are discussed at
the end. Note, this is also a phase diagram for the spin 1/2 XXZ
magnet, and the regime of proposed RVB phase of
Fazekas-Anderson \cite{FazekasAnderson} is actually a particular
spin ordered state.





\begin{figure}
\includegraphics[scale=0.8]{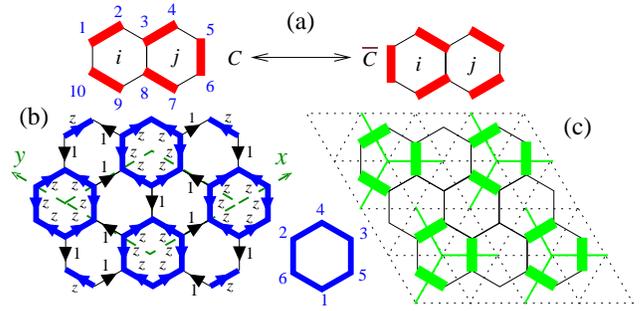}
 \caption{(Color online)
(a): two double-hexagon resonance configurations $c_{ij}$ and
$\bar{c}_{ij}=c_{ji}$. Red thick bars denote dimers.
(b): Kasteleyn orientation and edge weights of the honeycomb
lattice. Thick blue edges have weight $z$,  others have weight $1$.
The green dash-line rhombus encloses  the enlarged unit cell.
$x$,$y$ are the principal axis.  We use the six sites on a
thick-edge hexagon as the basis,  labeled as $1,\dots ,6$ as shown
in the right-bottom corner. (c): special edges (thick green on the
honeycomb) for the unitary transformation relating the unfrustrated
and frustrated case. Thin solid green bonds on the triangular
lattice are dual to the special edges. }
 \label{fig:doublehexagon}
\end{figure}

{\em Strong Repulsion Limit and Generalized Marshall Sign:} In the
limit of $V \gg |t|$, we can restrict the Hilbert space to a
manifold of states which correspond to classical Ising ground states
of the triangular antiferromagnet \cite{Wannier50}. Every such Ising
configurations $\ising$ can be represented by a close-packed dimer
configuration $\dimer$ on the dual honeycomb lattice. This is a
two-to-one mapping because of the Ising Z$_2$ symmetry
(particle-hole symmetry in the boson language).
The Hamiltonian (\ref{equ:H}), projected into this
degenerate subspace, introduces dynamics which splits the
degeneracy. Note, to first order in degenerate perturbation theory,
only the hopping term
$H_{t}=-t\sum_{\langle ij\rangle}(b^{\dag}_i b_j^{\vphantom{\dag}}+ H.c.)$
plays a role, leading to the double-hexagon
resonance in \fig{fig:doublehexagon}~(a) with amplitude $-t$. The
problem of the large repulsion limit is therefore related to finding
the ground state of a quantum dimer model with such dimer
resonances. We have already noted that the $t>0$ case is tractable
by Quantum Monte Carlo methods since there is no sign problem.
However, the problem of interest here is the case  $t<0$. {\em If}
there is a unitary transformation which changes the sign of every
matrix element of $H_{t}$, the problem can be mapped to unfrustrated
case.
This is generically not possible but, {\em within} the restricted
Hilbert space, this indeed happens, and the required unitary
transformation is the following.
Consider the lattice in \fig{fig:doublehexagon}~(c) with $1/4$
special edges marked as thick and green. One can check by inspection
that any double-hexagon resonance will change the number of covered
special edges by $\pm 2$. Therefore, if we define a unitary
transformation on the dimer basis \be
 |\dimer'\ket = U|\dimer\ket = \im^{N_s(\dimer)}|\dimer\ket,
 \label{equ:U}
\ee where $\im=\sqrt{-1}$, and $N_s(\dimer)$ is the number of special
(green) edges covered by a dimer in the dimer configuration
$\dimer$, the sign of the Hamiltonian will be changed. The unitary
transformation does not change the energy spectrum nor correlations
that are diagonal in boson density. Hence, thermodynamics -
that only depends on energy eigenvalues- is unchanged,
for eg. transition temperature and nature of transitions. However, off-diagonal
correlations are affected. We can therefore immediately conclude
that the ground state has the same three sublattice density
modulation as the supersolid phase in the unfrustrated model.
Moreover, it also has a finite compressibility, identical to that in
the unfrustrated problem, since this can also be expressed as a
density-density correlation function. The latter strongly suggests
superfluid long range order (a 2D bosonic phase with finite
compressibility at zero temperature), and taken all together this points
towards supersolid order for $t/V = 0^-$ as well. In order to
directly establish off diagonal long range order, and obtain more
detailed quantitative information, we turn to a variational
wavefuction approach.

{\em Variational Wavefunction} We denote the two
Ising states related to the dimer state $\dimer$ as
$\ising[\dimer]$ and $\bar{\ising}[\dimer]$ and consider the
following kind of wavefunctions,
\be
|\Psi\ket=\sum_{\dimer}\phi(\dimer)|\dimer\ket=\sum_{\dimer}\phi(\dimer)\cdot
  \left (|\ising[\dimer]\ket+|\bar{\ising}[\dimer]\ket \right )/\sqrt{2}
 \label{equ:psi}
\ee
where $\phi(\dimer)$  is the (complex) amplitude.
In the dimer representation, the projected $H_{t}$ corresponds  to
the double-hexagon resonance in \fig{fig:doublehexagon}~(a). Only
those dimer configurations with `resonatable' double hexagons appear
in the Hamiltonian matrix elements. We denote by $c_{ij}$, that a
particular dimer covering has a resonatable double hexagon at the
pair of adjacent plaquettes $i,\,j$, where $i$ is the plaquette with
two dimers. Under resonance $c_{ij}\rightarrow \bar{c}_{ij}=c_{ji}$.
However, the rest of the dimer configuration with this pair of
plaquettes removed $d_{ij}$ remains unchanged. Hence, the entire
dimer configuration may be denoted as $c_{ij}+d_{ij}$. Note, a
single dimer configuration may have many representations in this
notation - one for each resonatable hexagon pair.
The variational energy $E=\bra\Psi|H_{t}|\Psi\ket$ is
\be
 E= -t\sum_{<ij>}\sum_{d_{ij}}\left [ \phi^*(c_{ij}+d_{ij})
  \phi(\bar{c}_{ij}+d_{ij}) + c.c.\right ]
 \label{equ:E-as-phi}
\ee
where $c.c.$ is the complex conjugate. Before considering the frustrated case in detail, we briefly review
the variational wavefunction for the unfrustrated case\cite{Sen08}. There, $t > 0$, so
the matrix elements of $H_{t}$ are all non-positive in the dimer basis.
Thus the Perron-Frobenius theorem applies and the ground state can be taken to be everywhere positive.
Hence, we get a normalizable wavefunction if $\phi(\dimer)=\sqrt{P(\dimer)}$
with $P(\dimer)$ taken as the probability of the dimer configuration $\dimer$. Equivalently, one can assign positive weights $W(\dimer)$ to each dimer configuration $\dimer$, then the probability $P(\dimer)=W(\dimer)/Z$,  where $Z=\sum_{\dimer}W(\dimer)$.
\par
The central assumption that leads to tractable wavefunctions is the following. We assign edge weights $w_{ab}$ to all honeycomb lattice edges $ab$, and write the weight of dimer covering $\dimer$ as $W(\dimer)=\prod_{{\rm covered\ } \langle ab\rangle} w_{ab}$. Interpreting $W$ as a ficticious Gibbs weight, this corresponds to a problem of hardcore dimers in an external potential. Powerful  Grassmann variable techniques have been developed for this problem, which will allow us to calculate properties of these wavefunctions.

Plug the ansatz $\phi(\dimer)=\sqrt{P(\dimer)}$ into
(\ref{equ:E-as-phi}), and using the fact that the ratio $ \frac{ P(\bar{c}_{ij}+d_{ij}) }{
P(c_{ij}+d_{ij}) } $ is independent of the configuration $d_{ij}$.
\be
  E=-t\sum_{<ij>}\sqrt{P(c_{ij}) P(\bar{c}_{ij})}
  \label{equ:E}
\ee where  $P(c_{ij})=\sum_{d_{ij}}P(c_{ij}+d_{ij})$ is the net probability of the local configuration.


The dimer number operator is
 $n_{ab}=(1+s_i^z s_j^z)/2$
where $ab$ is the honeycomb lattice edge dual to the
triangular lattice edge $ij$.
The probability $P(c_{ij})$, $P(\bar{c}_{ij})$
are the expectation values $\bra n_{12}n_{34}n_{56}n_{78}n_{9,10}\ket$,
$\bra n_{23}n_{45}n_{67}n_{89}n_{10,1}\ket$, respectively.
This can be evaluated analytically by the Grassmannian
integral method \cite{Samuel}.
In the Grassmannian formulation, the dimer partition function is represented as
an integral over Grassmannian variables $\eta_a$ defined on
the honeycomb lattice sites, $Z=\int \exp(\sum_{a,b}\eta_a A_{ab} \eta_b/2) \prod_{a}\eta_a={\rm Pf}[A]$,
where
${\rm Pf}[A]$ is the Pfaffian of the Kasteleyn matrix $A$ \cite{Kasteleyn1963},
and $A_{ab}=+ w_{ab}$ if the Kasteleyn orientation is from $a$ to $b$,
or $=- w_{ab}$ if otherwise (see FIG.~\ref{fig:doublehexagon}~(b)).
The probability $P(c_{ij})$ is calculated as an expectation value in the Grassmannian theory,
the rule is to replace $n_{ab}$ by $A_{ab}\eta_a\eta_b$, then we get
$
 P(c_{ij})
  = w_{12}w_{34}w_{56}w_{78}w_{9,10}
  \left | \bra\prod_{a=1}^{10}\eta_a\ket \right |.
$
Thus the variational energy (\ref{equ:E}) can be written as
$
 E=-t\sum_{<ij>}{\sqrt{\prod_{i=1}^{10} w_{i,i+1} } }
\left |\bra\prod_{a=1}^{10}\eta_a\ket \right |
$
with $w_{10,11} =w_{10,1}$.
The ten-point correlator of anticommuting $\eta$ can be Wick-expanded into a
Pfaffian of a $10\times 10$ antisymmetric matrix,
$ \left | \bra\prod_{a=1}^{10}\eta_a\ket \right |
 = {\rm Pf}[\bra\eta_a\eta_b\ket]
 = \sqrt{\det[\bra\eta_a\eta_b\ket]},\ a,b=1\dots 10.
$
The above formula can be further simplified to the determinant of
a $5\times 5$ matrix exploiting the bipartiteness of the honeycomb lattice:
$
 \left | \bra\prod_{a=1}^{10}\eta_a\ket \right | =
  |\det[\bra\eta_a\eta_b\ket]|,\ a=1,3,\dots ,9;\ b=2,4,\dots ,10.
$.
{\modified This is much more efficient than the brutal-force Wick
expansion used by Sen {\it et al.} \cite{Sen08}, which allows us to
evaluate more complicated correlation functions later in this
paper.} 
The two-point correlator $\bra\eta_a\eta_b\ket=(A^{-1})_{ba}$
can now be evaluated by a Fourier transformation since the Kasteleyn
matrix $A$ has 2D translational symmetry. For the chosen Kasteleyn
orientation and basis shown in FIG.~\ref{fig:doublehexagon}~(b)),
in the thermodynamic limit, the two-point correlator of the site
$a$ in unit cell $(0,0)$ and the site $b$ in unit cell $(x,y)$ is
\bes
 \bra\eta_{a,(0,0)}\eta_{b,(x,y)}\ket=\int_{0}^{2\pi}\int_{0}^{2\pi}
  [\tilde{A}^{-1}(\vec{k})]_{ba} e^{\im(k_x x+k_y y)} \frac{\dif k_x \dif k_y}{4\pi^2}
\ees
where
$a,b=1,\dots,6$, and $\tilde{A}^{-1}(\vec{k})$ is the inverse of
the $6\times 6$ anti-hermitian matrix $\tilde{A}(\vec{k})$,
\bes
 \tilde{A}(\vec{k}) = \begin{pmatrix} 0_{3\times 3} & R(\vec{k}) \\
  -R^\dagger(\vec{k}) & 0_{3\times 3}
 \end{pmatrix},\,\text{with}\,
 R(\vec{k}) = \begin{pmatrix}
  \frac{1}{\epsilon_x \epsilon_y} & z & z \\
  z & \epsilon_y & z \\
  z & z & \epsilon_x
 \end{pmatrix}
\ees
where $\epsilon_x=e^{\im k_x},\,\epsilon_y=e^{\im k_y}$.
As is shown in \inlinecite{Sen08}, for $t>0$ this variational wavefunction has
two local minima at $z\approx 0.9258$ with energy per site $E=-0.13774t$,
and $z\approx 1.073$ with energy per site $E=-0.13762t$,
corresponding to the two supersolid states, $(+--)$ and $(0+-)$
of the triangular lattice boson model \cite{Wessel05, Heidarian05, Melko05}.
For the frustrated case, the wavefunction is obtained by unitary
transformation,
$|\Psi'\ket=U|\Psi\ket=\sum_{\dimer}\phi'(\dimer)|\dimer\ket$, hence
the variational wavefunction is
 $\phi'(\dimer)=\im^{N_s(\dimer)} \sqrt{P(\dimer)}$. The variational energy $E=\bra\Psi'|H_{t}|\Psi'\ket$ is of course the same as in the unfrustrated case.
In order to understand the two variational wavefunctions better,
we shall calculate two point correlation functions, assuming for simplicity that the two points  $i$, $j$ are on the same horizontal line,
{\modified and $j$ is on the right.}

{\em Diagonal Correlations:} Consider first the density-density
correlation function $\bra s^z_i s^z_j\ket$. Draw a line from $i$ to
$j$ and it will cut through an set of honeycomb lattice edges
$<ab>$. If the number of edges with no dimers cut by this line is
even, then $s^z_i s^z_j = +1$ and otherwise $= -1$. In terms of the
dimer number operator $n_{ab}$ the $s^z_i s^z_j$ becomes a non-local
string operator,
\be
  \bra s^z_i s^z_j \ket
= \bra \prod_{<ab>{\rm\ cut\ by\ }ij}(2n_{ab}-1)\ket
\ee
Expand the product we get $2^{|j-i|}$ terms($|j-i|$ is the distance between
 $j$ and $i$ measured by the triangular lattice constant),
each of which is the type of correlation functions evaluated before.
Because these operators are diagonal in the dimer basis,
$\bra\Psi|s^z_i s^z_j|\Psi\ket = \bra\Psi'|s^z_i s^z_j|\Psi'\ket$.

\begin{figure}
 \includegraphics{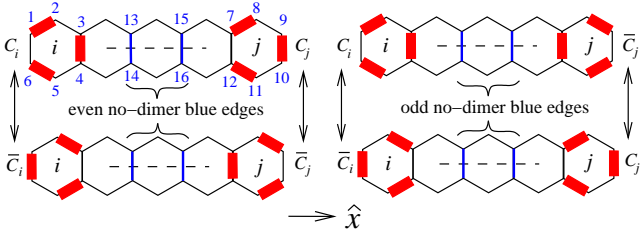}
 \caption{(Color online)Two possible simultaneous double-resonance
  needed for calculating $\bra b^\dagger_i b_j\ket$:
  $\dimer_i,\dimer_j \leftrightarrow \bar{\dimer}_i,\bar{\dimer}_j$,
  and $\dimer_i,\bar{\dimer}_j \leftrightarrow \bar{\dimer}_i,\dimer_j$,
  with even(odd) number of no-dimer edges cut by
  the line $i+\hat{x},j-\hat{x}$(dash line).
 }
 \label{fig:doubleresonance}
\end{figure}

{\em Off diagonal Correlations:}
The square of the off-diagonal long range order (ODLRO) parameter
 $\bra b^\dagger_i b^\nd_j \ket$ is slightly more complicated.
In the dimer basis it describes the simultaneous resonances of two hexagons
(if $i$ and $j$ are not neighbors).
However for this process to happen $s^z_i$ and $s^z_j$ must be opposite.
Label the two local resonating configurations on hexagon $i$($j$) by
 $\dimer_{i(j)}$and $\bar{\dimer}_{i(j)}$,
 there are two possibilities of this simultaneous double-resonance,
 shown in \fig{fig:doubleresonance},
 with opposite conditions for the edges cut by the line $i+\hat{x},j-\hat{x}$,
 where $\hat{x}$ is the horizontal triangular lattice vector.
The even(odd) requirement can be enforced using the dimer number operators as
 $ [1\pm\prod(2n_{ab}-1)]/2 $, where the product is over all edges
 $<ab>$ cut by the line $i+\hat{x},j-\hat{x}$ (see \fig{fig:doubleresonance} for an example).
Consider $t>0$ case first, we have
\be
 \begin{split}
  & \bra\Psi|b^\dagger_i b^\nd_j|\Psi\ket =
   \frac{ w_{23}w_{45}w_{61} w_{89}w_{10,11}w_{12,7} }
    { w_{12}w_{34}w_{56} w_{78}w_{9,10}w_{11,12} } \\
  & \ \times \Big \{
   \left \bra n_{12}n_{34}n_{56} n_{78}n_{9,10}n_{11,12}\cdot
    [ 1+\prod(2n_{ab}-1) ]/{ 2 } \right \ket \\
  & \ \ \ + \left \bra n_{12}n_{34}n_{56} n_{78}n_{9,10}n_{11,12}\cdot
    [ 1-\prod(2n_{ab}-1) ]/{ 2 } \right \ket
   \Big \} \\
  & = \sqrt{ \prod w } \left |\bra \prod_{a=1}^{12}\eta_a\ket \right |
 \end{split}
 \label{equ:spsm}
\ee
where the
$\prod w$ is the product of edge weights of the twelve(12) edges around
 hexagons $i$ and $j$. Note, this simple form arises because the string $\prod(2n_{ab}-1)$ cancels out. We will see shortly that in the frustrated hopping case, this does not happen.

If distance between $i$ and $j$ is large, the 12-point correlator
 $\bra\prod_{a=1}^{12}\eta_a\ket$ can be factorized into two 6-point
correlators
 $\bra\prod_{a=1}^{6}\eta_a\ket\cdot\bra\prod_{a=7}^{12}\eta_a\ket$.
And we have the relation
 $\sqrt{ w_{12}w_{34}w_{56} w_{23}w_{45}w_{61} }|\bra
 \prod_{a=1}^{6}\eta_a\ket| = \bra\Psi|b^\dagger_i|\Psi\ket $.
So literally we have the factorization property
\bes
 \bra\Psi| b^\dagger_i b^\nd_j|\Psi\ket \to
 \bra\Psi | b^\dagger_i|\Psi\ket \bra\Psi| b^\nd_j|\Psi\ket,\quad
 |j-i| \to \infty
\ees


For $t < 0$ case we need to take care of the phases of $\phi'$.
From \fig{fig:doublehexagon}~(c) we can see that $\bra\Psi'|b^\dagger_i b^\nd_j|\Psi'\ket$
has similar form as the first line of \eqtn{equ:spsm}, only the first term
inside $\big \{ \cdot\big \}$ acquires a minus sign. Therefore we get
\be
 \begin{split}
  & \bra\Psi'|b^\dagger_i b^\nd_j|\Psi'\ket =
   -\frac{ w_{23}w_{45}w_{61} w_{89}w_{10,11}w_{12,7} }
   { w_{12}w_{34}w_{56} w_{78}w_{9,10}w_{11,12} } \\
  & \quad \times \left \bra n_{12}n_{34}n_{56} n_{78}n_{9,10}n_{11,12}
   \prod(2n_{ab}-1) \right\ket
 \end{split}
\ee
The product can be expanded into $2^{|j-i|-2}$ terms,
each of which can be evaluated as before.
Note, this correlation function cannot be factorized as in
the unfrustrated case and one necessarily needs to evaluate a string correlator.

{\it Results:}
We evaluate the above mentioned correlators up to distance $|j-i|=18$ and
extrapolate to infinite distance limit to determine the long range order.

At the global energy minimum $z=0.9258$ and for the {\em
unfrustrated} ($t>0$) case, the long range order parameter $\bra
\vec{s}\ket=(b^\dagger+b^\nd,\im b^\nd-\im b^\dagger,2n-1)=
(0.163,0,0.764),\ (0.372,0,-0.412),\ (0.372,0,-0.412)$ for the three
sublattices A,B,C, respectively (we have set the superfluid phase to
zero and sublattice A is surrounded by weight $z$ hexagon in
\fig{fig:doublehexagon}~(b)). These numbers are in agreement with
Quantum Monte Carlo (QMC) results. The average density deviation
from 1/2 is $|0.764-0.412-0.412|/2/3=0.010$, which is about $2\%$,
in good agreement with QMC \cite{Heidarian05}. The solid order
parameter is $ |n_A+n_B e^{2\pi\im/3}+n_C e^{4\pi\im/3}|^2/9 =
0.0384$ ($n_{A,B,C}$ are boson densities on the three sublattices),
which is about $15\%$ smaller than the QMC result of $0.045$
\cite{Heidarian05, Prokofiev}, but in good agreement with classical
Monte Carlo calculations result $0.0389$ of the same type of
variational wavefunctions \cite{Sen08}.

In the {\em frustrated} ($t<0$) case the three sublattice order is
$(0,0,0.764),\ (0.389,0,-0.412),\ (-0.389,0,-0.412)$, as shown in
Fig. \fig{fig:order}. The average density deviates from 1/2 by the
same amount as the frustrated case. In spin language this means a
non-zero average z-component of spin,
$|\sum_{i}S^z_i/N|=|\sum_{i}s^z_i/(2N)|=0.01$, which is about $50\%$
smaller than harmonic spin-wave result $0.02$, which has the same
symmetry\cite{Fazekas}. Note that surprisingly the superfluid
amplitude($XY$-component of $\vec{s}$) on the B,C sublattices is
{\em larger} than those in the unfrustrated case, while it vanishes
on the A sublattice. Note, this quantity can be directly measured in
Quantum Monte Carlo simulations of the unfrustrated system in the
large repulsion limit, by calculating correlations of the unitarily
transformed operator. For example, with $O=s^+_i s^-_j$,with $j$ to
the right of $i$ in the same horizontal line and $|j-i| > 2$, the
correlator to be measured is $U^\dagger O^\nd U^\nd=-s^+_i
s^z_{i+\hat{x}} s^z_{j-\hat{x}} s^-_j$. Finally, we combine the
present results in the large repulsion (or $V\gg -t$) limit with
known $120^\circ$ order in the isotropic $V=-2t$ ($J_z=J_\perp$)
limit. These can be connected without a phase transition, is we
assume that the supersolid phase persists with no change in symmetry
over the entire range $V>-2t>0$. This scenario, which is also a
phase diagram for the spin 1/2 XXZ antiferromagnet, is depicted in
Figure \ref{fig:phasediagramXXZ}. \footnote{ At the other local
minimum $z=1.073$, the three sublattice order for unfrustrated and
frustrated cases are $(0.438,0,0),\ (0.231,0,0.661),\
(0.231,0,-0.661)$ and $(0.340,0,0),\ (-0.147,0,0.661),\
(-0.147,0,-0.661)$.}


{\it Experimental Realization:} How can the frustrated boson
hoppings be experimentally realized? In lattice cold atom systems, a
recent experiment \cite{Winkler} demonstrated that `repulsively'
bound molecular bosons have frustrated hoppings. Consider preparing
an initial state composed of molecules of pairs of atoms (either
bosons or fermions) with one or zero molecules per site. If the
interactions between atoms are now made repulsive, the effective
molecular hopping is readily seen to be frustrated, since the singly
occupied sites are lower in energy. If this metastable state is
sufficiently long lived, the equilibrium properties of the
frustrated boson system can be accessed. In Josephson Junction
Arrays, external magnetic fields can generate frustrated
hopping\cite{Chaikin}.

We acknowledge funding from NSF-DMR 0645691 and ARO Grant No.
W911NF-07-1-0576, and useful discussions with Ehud Altman and Arnab Sen.


\begin{acknowledgments}
\end{acknowledgments}

\end{document}